\documentclass[sn-aps]{sn-jnl}
\theoremstyle{thmstyleone}

\theoremstyle{thmstyletwo}

\theoremstyle{thmstylethree}

\begin{document}

\title{Fragile electronic superconductivity in Bi Single crystal}
\author[1]{\fnm{Anil} \sur{Kumar}}
\author[1]{\fnm{Rajendra} \sur{Loke}}
\author[2]{\fnm{Arindam} \sur{Pramanik}}
\author[2]{\fnm{Rajdeep} \sur{Sensarma}}
\author[3]{\fnm{Sitaram} \sur{Ramakrishnan}}
\author*[4]{\fnm{Om} \sur{Prakash}}\email{om.prakash.shukla@icloud.com}
\author[5]{\fnm{Biplab} \sur{Bag}}
\author[1]{\fnm{Arumugam} \sur{Thamizhavel}}
\author*[6]{\fnm{Srinivasan} \sur{Ramakrishnan}}\email{ramky07@gmail.com}
\affil[1]{\orgdiv{Department of Condensed Matter Physics and Materials Science},
\orgname{Tata Institute of Fundamental Research}, \orgaddress{\city{Mumbai}, \postcode{400005}, \country{India}}}
\affil[2]{\orgdiv{Department of Theoretical Physics},
\orgname{Tata Institute of Fundamental Research}, \orgaddress{\city{Mumbai}, \postcode{400005}, \country{India}}}
\affil[3]{\orgdiv{Department of Quantum Matter, AdSE},
\orgname{Hiroshima University}, \orgaddress{\city{Higashi-Hiroshima}, \postcode{739-8530}, \country{Japan}}}
\affil[4]{\orgdiv{Quantum Computing Hardware},
\orgname{VTT Technical Research Centre of Finland}, \orgaddress{\city{Espoo}, \postcode{02150}, \country{Finland}}}
\affil[5]{\orgdiv{Amity Institute of Applied Sciences},
\orgname{Amity University Jharkhand}, \orgaddress{\city{Ranchi}, \postcode{834002}, \country{India}}}
\affil[6]{\orgdiv{Department of Physics},
\orgname{Indian Institute of Science Education and Research}, \orgaddress{\city{Pune}, \postcode{411008}, \country{India}}}

\abstract{It was presumed that semimetal Bismuth (Bi) would not show superconductivity (SC) even at ultra-low temperatures ($<$10 mK)
due to its very low carrier density ($\approx 3\times10^{17}$cm$^{-3}$). Recently, we have established bulk superconductivity in
ultra-pure (99.9999\%) Bi single crystal at $\mathrm{T_C = 0.53}$ mK with an extrapolated upper critical field $\mathrm{H_C(0) = 5.2\mu}$T measured along the [$0001$] (trigonal) -crystallographic direction \cite{prakash2017a}. At very low concentrations of the charge carriers,
we are dealing with fragile Cooper pairs with an estimated large coherence length $\mathrm{\xi_{GL}(0)\approx 96 \mu}$m.
We also stated that one needs to go beyond the conventional electron-phonon coupling (BCS-like) mechanism to understand the SC state in Bi. Bi is a compensated semi-metal with electrons and holes as charge carriers. In order to find the charge carriers responsible for the SC, we report the temperature dependence of the
anisotropic critical field along the [$01\bar 10$] (bisectrix)-crystallographic direction and compared it with the earlier data from measurements along the trigonal.
Our theoretical analysis of the anisotropy of critical fields suggests that the 
light electrons in the three pockets of Bi bands are responsible for the SC and  indicates that
Bi is an extremely weak type-II (close to type-I) superconductor. Finally, we review the current theories proposed to explain the SC in Bi.}




\maketitle

Although semimetal Bi has been studied for more than a century, it still draws
attention from both theorists and experimentalists who work in the frontier areas of
condensed matter research \cite{edelman1976a}. Bismuth with the electronic configuration ([Xe]- 4f$^{15}$ 5d$^{10}$ 6s${^2}$ 6p${^3}$)
crystallizes into a distorted rhombohedral structure described by the space  group $R\bar{3}m$ (166) with $\bf{a}$ $=$ 4.538 \AA ~and $\bf{c}$ $=$ 11.823 \AA.
The unit cell consists of two pentavalent Bi atoms, giving rise to a Fermi surface (FS) comprising of four ellipsoidal-shaped pockets: three electron and one hole pocket; making Bi nearly charge compensated semimetal.
These pockets account for a very small ($\approx 10^{-5}$) area of the FS resulting in a carrier density of $\mathrm{n \approx p \approx 3\times 10^{17}/cm^3}$, small density of states at FS $\mathrm{g(E_F)\approx 4.2 \times 10^{–6}/(eV-atom)}$ (see figures 1-3 in SM). The low carrier density translates to a single charge carrier shared by nearly 100,000 Bi atoms. This led to a presumption that, if at all, superconductivity in bulk Bi would occur only at ultra-low temperatures. Indeed in 2016 \cite{prakash2017a} studying a high purity (6N) single crystal of Bi, some of us established superconductivity below $\mathrm{T_{C} = 0.53 mK}$ with an estimated upper critical field of $\mathrm{H_C(0)\approx 5.16 \pm 0.06\mu T}$ along the [$0001$] (trigonal)-crystallographic direction.

Extensive studies on the charge carriers have established that the electrons in Bi have small effective masses ($\mathrm{m_{eff}\approx 0.001m_e}$) with significant anisotropy ($\mathrm{m_{eff}(B \parallel binary)/m_{eff}(B \parallel bisectix)\geq 200}$), while the holes are heavier ($\mathrm{m_{eff}\approx 0.07m_e}$) and have relatively isotropic effective mass (anisotropy $\approx 10$)\cite{Zhu2011}. To better understand the nature and possible mechanism of superconductivity in Bi, we have studied the critical field anisotropy in superconducting Bi using dc magnetization measurements. Here, we report measurement of the temperature dependence of the upper critical field in a 6N highly pure single crystal of Bi along the [$01\bar 10$] (bisectrix)-crystallographic direction and present a comparison with the critical field measured along the trigonal \cite{prakash2017a}. Our theoretical analysis of the electronic g-factor, together with the observed critical field anisotropy suggest
that the light electrons are primarily responsible for the superconductivity in Bi.

\begin{figure}[H]
\centering
\includegraphics[width=\textwidth]{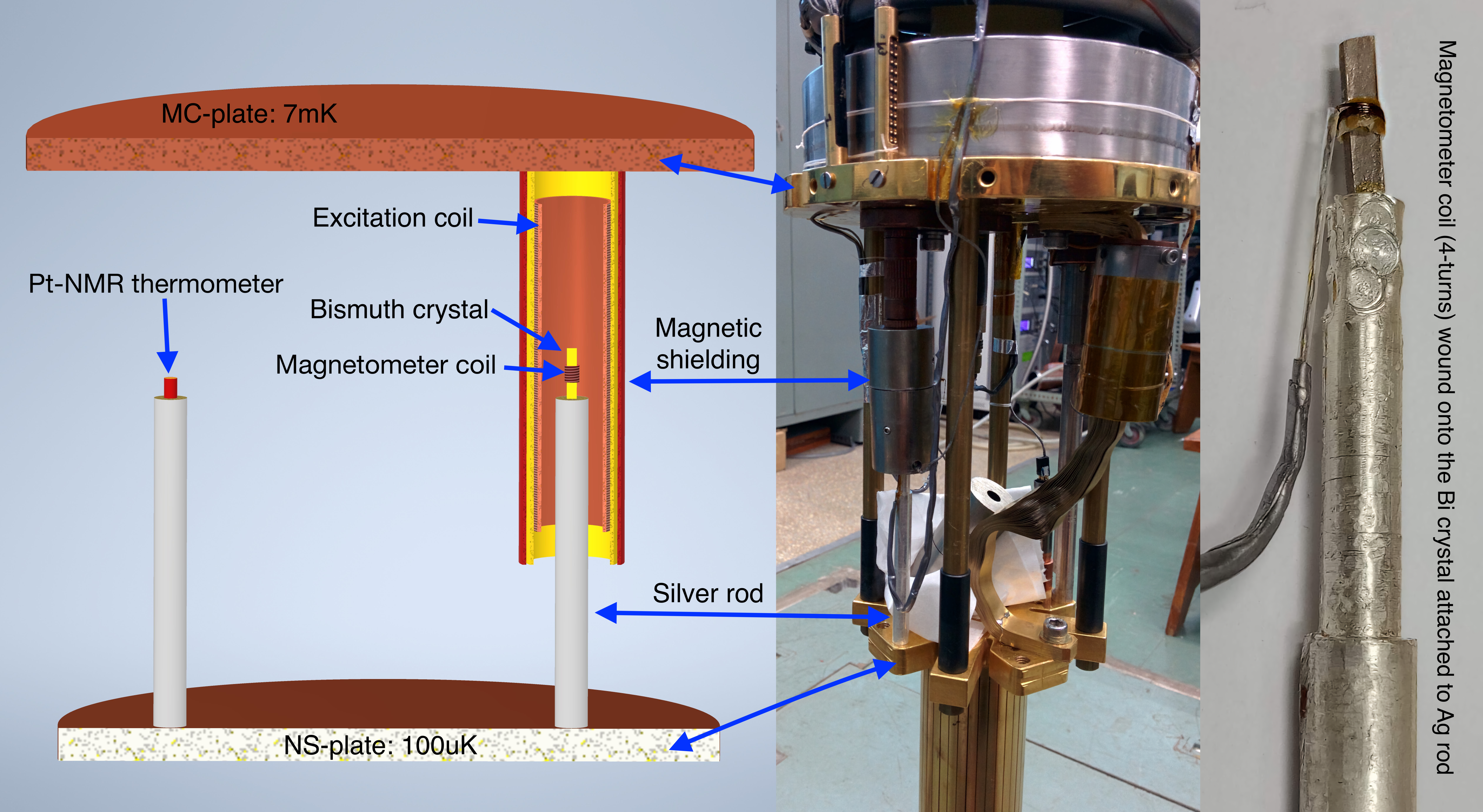}
\caption{The schematic drawing of the measurement setup with magnetic shielding assembly: (Right) The overview of the measurement setup consisting of an
excitation coil ($\mathrm{B/I\approx 0.04\mu T/\mu A}$) enclosed in magnetic shields attached to the MC, Pt-NMR thermometer and samples mounted on the NS plate (100 $\mu$K). (Middle) Actual measurement setup in the dilution insert. The arrows mark different components shown in the drawing.
(Left) Bi single crystal oriented along the [$01\bar 10$] (bisectrix)-crystallographic direction is attached to the silver rod. The measurement setup used here is similar to the thermalization and measurement arrangements reported earlier \cite{prakash2017a}.}
 \label{fig1}
\end{figure}

To study the upper critical field along bisectrix, a Bi crystal sized  $2.1\times 0.3 \times 0.2$ cm$^{3}$ (from the same batch used in \cite{prakash2017a}, RRR $\geq 500$) was cut along the [$01\bar 10$]  crystallographic direction (see figure 1 in SM). The crystal was
push-fitted to an annealed high-purity (5N) silver (Ag) rod with fine Ag powder at the interface to enhance effective surface contact area (see fig. [\ref{fig1}]). The Ag rod was further pressed onto the crystal to provide better thermal contact and threaded to the Cu nuclear stage (NS), enabling cooling of the sample down to $\mathrm{T\approx 100 \mu}$K. The dc-magnetization measurement assembly consists of a magnetometer pick-up coil (4 turns) and an excitation coil, both made up
of superconducting niobium (Nb). The pickup coil as well the excitation coil outgoing leads are twisted in pairs and enclosed in Pb foil to minimise coupling to the external magnetic fields and reduce noise in the measurements. The measurement assembly was enclosed in a magnetic shielding consisting of high permeability material Cryoperm-10 ($\mu_r \geq 10^4$ at 4.2K, SEKELS GmbH, Germany) and superconducting lead (Pb) shields and attached to the mixing chamber (MC, T$=7$mK) plate of the dilution fridge (DRS 1000, Leiden
Cryogenics, Netherlands). This magnetic shielding setup is tested to shield the sample from the
external magnetic fields of the order of $\mathrm{B_{ext}\approx 10 mT}$ down to $\mathrm{B_{shielded} \lesssim 10 nT}$, when there is no current in the primary coil. The magnetometer coil is wound directly onto the single crystal to maximize the filling fraction and connected to the input coil of the dc-SQUID (Tristan Technologies, USA). Figure [\ref{fig1}]
shows the schematic drawing of the measurement setup.

Apart from shielding external magnetic fields, the magnetic shields also decrease the field produced by the enclosed primary coil. The excitation coil was calibrated while enclosed in the magnetic shielding assembly at 4.2 K using a
single-axis magnetometer with the low field probe (Bartington Instruments Ltd, England, $\mathrm{\pm 1nT}$ resolution) to accurately determine the excitation magnetic fields used in the measurements. The pick-up coil is connected to the dc-SQUID (Fig [\ref{fig1}]),
which in turn are connected to the RF-amplifier fixed at the head of the cryostat at room temperature.
The RF-head is connected to the squid control unit which directly reads output in volts. The dc$-$SQUID
output has been calibrated at 4.2 K by measuring the diamagnetic signal from classical superconductors,
Nb and Pb. To calibrate the SQUID output voltage with
the diamagnetic susceptibility, we used a Rh metal of the same dimension
as Bi crystal and measured the jump in the SQUID output voltage at the transition temperature with different
excitation fields. We used similar excitation and pick-up coil setups consisting of the magnetic
shields mentioned above for calibration (see \cite{prakash2017a}).

The experiment was carried out in a dilution refrigerator equipped with a single Cu-adiabatic nuclear demagnetization stage.
The Cu-nuclear stage was first cooled by the dilution refrigerator down to 7 mK followed by magnetization
of the Cu-nuclear spins by applying a magnetic field of 9 T using a superconducting magnet (Cryogenics, UK).
The Cu-nuclear stage was thermally connected to the mixing chamber using an Aluminium (Al) superconducting thermal
switch to facilitate isothermal magnetization. The application of the 9 T magnetic field heats up the Cu-stage
to nearly 40 mK due to the heat of magnetization and we have to wait for nearly 36 hours to cool down the
magnetized Cu-stage to 10 mK. Subsequently, the Al thermal switch is turned off to thermally disconnect the
Cu-NS from the MC by removing the current in the solenoid enclosing Al switch. A slow
adiabatic demagnetization of Cu nuclear spins over a period of 48 hours cools down the NS to a base
temperature of 100 $\mu$K. Slow demagnetization helps in maintaining thermal equilibrium between the samples and
NS as well as with the NMR thermometer. We used $^{195}$Pt-NMR thermometer for the temperature measurements
below 10 mK during adiabatic demagnetization. The NMR thermometer is calibrated against a Cerium Magnesium
Nitrate (paramagnetic thermometer) and a SQUID based noise thermometer (MAGNICON GmbH, Germany) at 10 mK.
The SQUID based noise thermometer can also measure temperatures down to 1mK and is used along with the NMR
thermometer below 10 mK. The details of the adiabatic TIFR nuclear refrigerator with temperature
measurement and calibration are given in an earlier report \cite{naren201a}.

\begin{figure}[H]
\centering
\includegraphics[width=\textwidth]{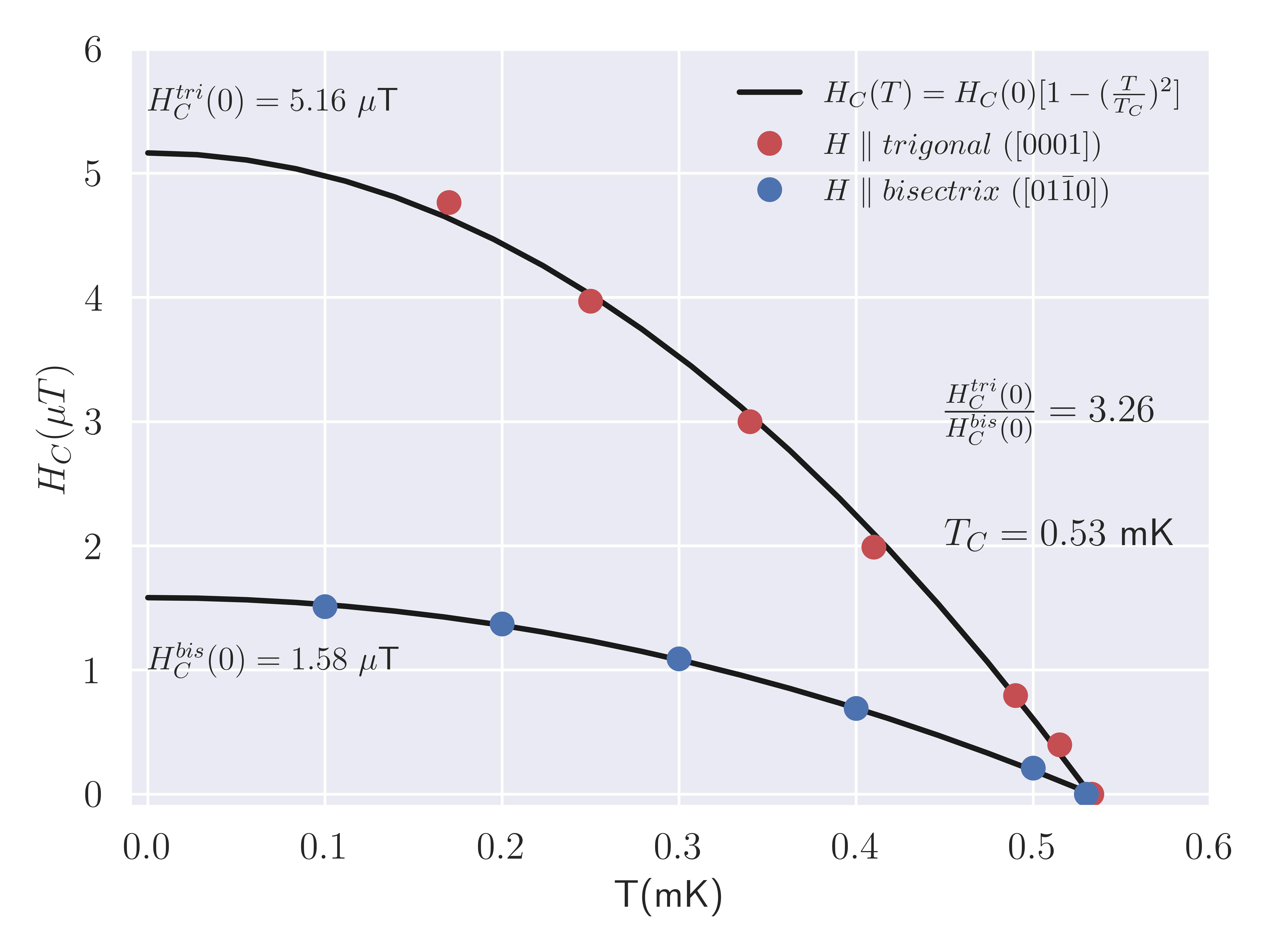}
\caption{The temperature dependence of the critical fields $\mathrm{H_C}$ for field along the trigonal and bisectrix-crystallographic directions. }
 \label{fig2}
\end{figure}

The superconducting transition of the Bi in a excitation field of 0.2 $\mu$T along the bisectrix is observed (see figure 6 in SM) below 0.53 mK in the form of a sharp jump in the dc-susceptibility ($\mathrm{\chi}$). The $\mathrm{T_C = 0.53}$ mK is in agreement with the $\mathrm{T_C}$ observed for field along the trigonal  \cite{prakash2017a}. The SC transition is suppressed to lower temperatures with increasing excitation field.  All measurements were performed on the free end of the sample, away from the Ag rod joint, avoiding any artefact due to the interface effects. The temperature dependence of the critical fields along the trigonal and bisectrix is shown in fig.[\ref{fig2}].
The data fitted to $\mathrm{H_{C}(T) = H_{C}(0)[1-(T/T_C)^{2}}$], to estimate the value
of the critical field at T$=0$ K. The critical field value along the bisectrix estimated from
the fit is $1.58 \pm 0.07\mu$T. The resulting critical field anisotropy is:  $\mathrm{H_C^{tri}(0)/H_C^{bis}(0) = 3.26\pm 0.18}$.
This critical field anisotropy presents both a challenge and possible window into understanding the mechanism behind the suppression of superconductivity due to the magnetic field in Bi. If Bi was a strong type II-SC, the anisotropy can be explained in terms of the anisotropic superfluid stiffness of the material (inherited
from the anisotropic band dispersion of Bi). However, the presence of a sharp jump in the magnetization, and nearly identical diamagnetic susceptibility in the field cooled (FC) and zero field cooled (ZFC) \cite{prakash2017a}
precludes the possibility that Bi is a strong type II-SC. This is not strange given that majority of elemental superconductors are type-I superconductors.
If we consider Bi as a type-I superconductor, the critical field $\mathrm{H_C}$, is determined
by equating the free energy of the superconductor at zero magnetic field (field does not
penetrate below $\mathrm{H_C}$) with the free energy of the normal state at $\mathrm{H_C}$(0). The anisotropy can
then only arise from the dependence of the free energy of the normal state on the direction
of the applied field. This free energy is given by the equation $\mathrm{H^{2}_C/2\mu_0(1+\chi_v}$), where $\mathrm{\chi_v}$ is the
volume magnetic susceptibility of the material. For a strongly diamagnetic material
like Bi, $\mathrm{\chi_v\approx 10^{-4}}$ \cite{Fukuyama1970} the anisotropic effects from $\chi$$_{v}$ can be neglected. Thus,
it is hard to explain an anisotropic factor of 3.26 in terms of $\mathrm{H_C}$. Additionally, a simple estimate of $\mathrm{H_C}$, assuming an
energy gap $\Delta(0) = 1.764 \mathrm{k_BT_C}$ \cite{prakash2017a} with $\mathrm{T_C = 0.53}$ mK and pairing of electrons with the density of states
(per pocket per spin) at the Fermi level $N^e_0 = 9.2\times 10^{18}$ eV$^{-1}$ cm$^{-3}$ \cite{ruhman2017a} leads to an isotropic $\mathrm{H_C(0)=0.2\mu}$T which is much lower than the values observed experimentally. A similarly low value of $\mathrm{H_C}$(0) is obtained if one assumes
hole pairing instead of electron pairing. If both electrons and holes are paired, the estimated $\mathrm{H_C(0) = 0.3 \mu}$T
is still lower than the experimentally measured values.

\begin{figure}[H]
\centering
\includegraphics[width=\textwidth]{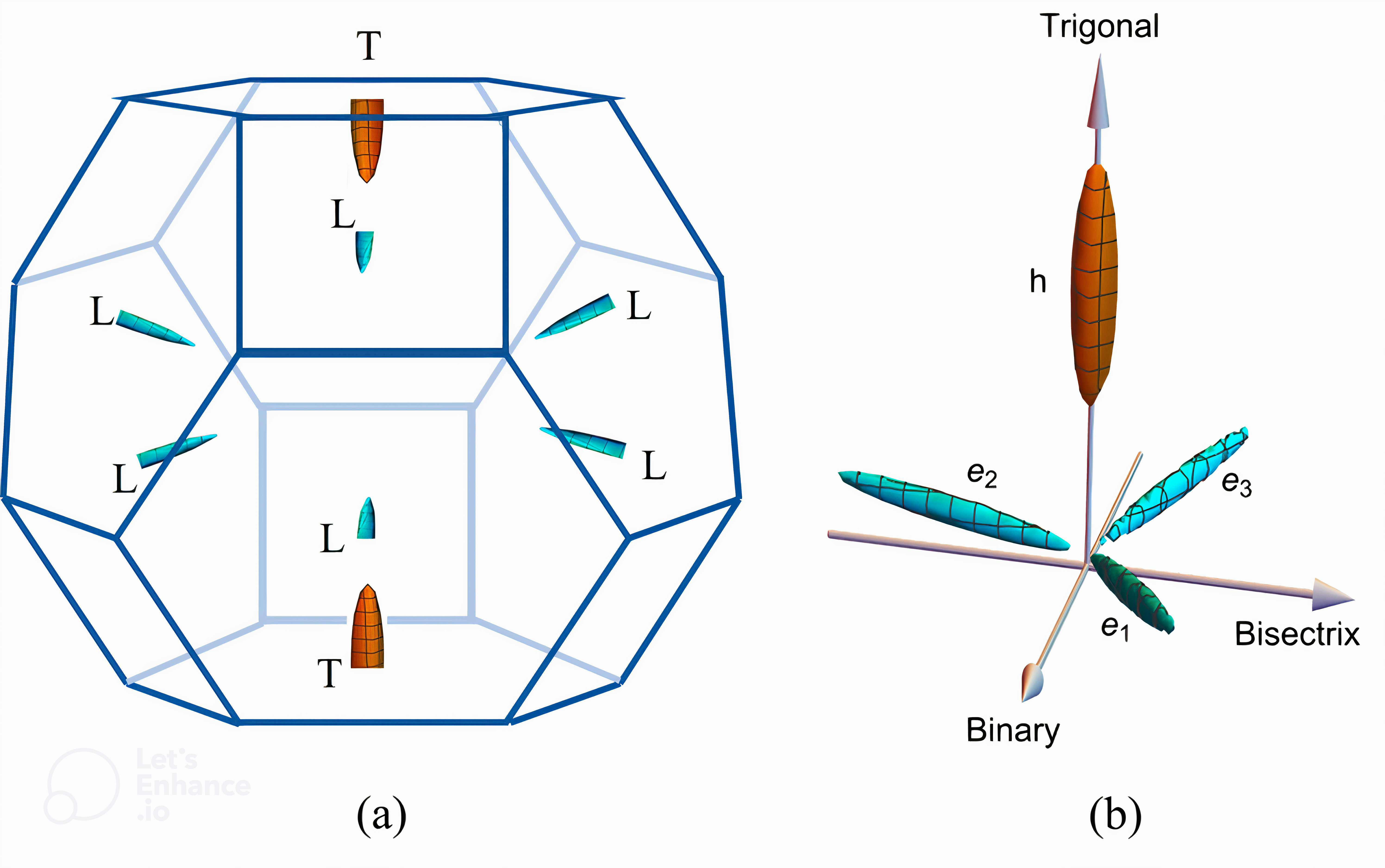}
\caption{(a) The first Brillouin zone and the Fermi surface of bismuth. (b) Enlarged electron (blue) and hole (red) pockets. The electron pockets make $\alpha \approx 7^{\circ}$ tilt angle with respect to binary-bisectrix plane. The pocket e2 lies along the bisectrix direction.}
\label{fig3}
\end{figure}

The anisotropy of the critical field can be understood if one assumes that Bi is a weakly type II superconductor
and the transition is given by the Pauli paramagnetic limit. In this case, one needs to compare the pairing energy
gain to the Zeeman energy cost paid by the electrons/holes to maintain similar Fermi surfaces for the two spin
components: $g\mu_B\mathrm{H_C}=\frac{\Delta(0)}{\sqrt{2}}$\cite{tinkham1996}. To further test this, we use the following model hamiltonian for each of the three electron pockets to extract the anisotropic $g$ factor.
\begin{equation}
H_e(\vec{k})=[\hbar v_zk_z\sigma^z+\hbar v_\perp(k_x\sigma^x+k_y\sigma^y)]s^x+\Delta_{bg}s^z
\end{equation}

where $\sigma^i$s and $s^i$s are the Pauli matrices in the spin and orbital space, respectively. We use
$v_z=6.6\times 10^4\, m/s$, $v_\perp=8.1\times 10^5 \,m/s$, and $\Delta_{bg}=7.5\, meV$ \cite{ruhman2017a}.
Here z denotes the direction along the $\mathrm{\Gamma-L}$ line in the BZ, which is also the long-axis of the electron pockets. Figure~[\ref{fig3}] shows the BZ and the
electron \& holes pockets contributing to the FS of Bi. The electron pockets have a tilt angle of $7^{\circ}$ with respect to the binary-bisectrix plane.

To find the spin-splitting in the presence of an external magnetic field, we block diagonalize
\cite{wolff1964, zhu2011} the hamiltonian in the orbital space to get the effective spin hamiltonian:\\
\begin{equation}
	H_{spin}=\frac{\hbar e}{\Delta_{bg}} v_zv_\perp H_x\sigma^x+\frac{\hbar e}{\Delta_{bg}} v_zv_\perp H_y\sigma^y+\frac{\hbar e}{\Delta_{bg}} v_\perp^2 H_z\sigma^z
\end{equation}\\
where $\textbf{H}$ is the external magnetic field. This allows us to extract the effective
$g$ factors for different electron pockets for $\textbf{H}$ in three different directions; trigonal,
binary, and bisectrix (see fig.[\ref{fig3}](b)), which are listed in table~[\ref{gfactor}]. These estimates (see SM for more details)
are roughly consistent with estimates from cyclotron resonances \cite{Zhu2011}.

\begin{table}[h]
\centering
\caption {Calculated $g$ factor anisotropy of the three electron pockets in Bi}
 \begin{tabular}{cccc}
 \toprule
 pocket & trigonal & binary & bisectrix \\
\midrule
 e$_1$ & 293 & 1716 & 1000 \\
 e$_2$ & 293 & 162 & 1979 \\
 e$_3$ & 293 & 1716 & 1000 \\
 \botrule
 \end{tabular}
 \label{gfactor}
\end{table}

We note that a larger value of $g$ in a particular direction implies a smaller critical field $\mathrm{H_C}$(0)
in that direction. From the table~[\ref{gfactor}], we see that the $g$ factor along the bisectrix direction is larger
than the $g$ factor along the trigonal direction, and hence one would expect the critical field along
 trigonal to be larger than the in-plane critical field along the bisectrix, as seen in our experiments. Further, the ratio of
$\mathrm{H_C}$(0) along trigonal to that along the bisectrix is $\mathrm{H_C^{tri}(0)/H_C^{bis}(0)=g_{bis}/g_{tri} \approx 3.41}$ (see table~[\ref{gfactor}]).
This matches quantitatively with the anisotropy ratio of $\approx 3.26$ measured experimentally. Our individual estimates for
$\mathrm{H_C^{tri}(0) \approx 3.4 \mu}$T and $\mathrm{H_C^{bis}(0) \approx 1\mu}$T are also close to the experimentally observed values. We also note that the $g$ factor
for the hole pocket is $\approx 63$ in the trigonal direction and $\approx 0.8$ in the plane \cite{Zhu2011}. Thus a Pauli limit
calculation with the hole pocket would yield a larger in-plane upper critical field compared to the critical field along the trigonal, contrary to the experimental findings. Thus the 
anisotropy seen in the experiments strongly points to the pairing of electrons driving superconductivity in Bi.

We now consider the models put forth to explain the observation of SC in Bi.
Earlier, Srinivasan, $et\hspace{0.1cm}al$. \cite{srinivasan1979a} used a simplified two-band model for describing
the longitudinal dielectric function and found that the attractive interaction responsible
for the instability of the normal ground state comes not only from the exchange of lattice phonons, but also from the
electron-hole sound mode, provided the ratio of the averaged hole to the electron mass, $\mathrm{m_h/m_e\neq 1}$.
This condition is easily satisfied in the case of Bi and they estimated the $\mathrm{T_C}$ to be around 1 mK. However, so far there is no experimental evidence for existence of the electron-hole sound 
(EHS) mode (a sound-like longitudinal collective mode) in Bi. Another model proposed by Zaahel Mata-Pinzón
$et\hspace{0.1cm}al$ \cite{pinzon2016a} numerically calculated electronic and
vibrational densities of states (eDOS and vDOS, respectively) of crystalline and amorphous forms of Bi.
The calculations showed that the eDOS of amorphous Bi (a-Bi)
is about $4 \times$ larger than that of crystalline Bi (c-Bi) at the Fermi energy, whereas, for the vDOS,
the energy range of the amorphous is roughly the same as the crystalline even though the actual
shapes are quite different. Using the experimental parameters obtained for a-Bi and
employing a simple weak coupling BCS mode, they gave an upper limit of 1.3 mK as the superconducting
transition for a pure Bi crystal. Furthermore, they suggested that the electron-phonon coupling
($\lambda$) is larger in a-Bi as compared to c-Bi as indicated by the $\lambda$
obtained via McMillan’s formula, $\lambda_c$ = 0.24 for c-Bi and while experiment and theory suggest $\lambda_a$ = 2.46 for
a-Bi. Therefore, with respect to the c-Bi, superconductivity in a-Bi is enhanced by the higher values of
$\lambda$ and eDOS at the Fermi energy. Though this model is simple and predicts $\mathrm{T_C}$ of the Bi
crystal near to its actual value, use of the electron-phonon mechanism in the formation of
Cooper pairs is not justified due to the failure of the  Migdal's adiabatic approximation \cite{migdal1958a}
which is central to any electron-phonon coupling scheme including the BCS-theory. This is due to the fact that the Fermi energy (25 meV)
and lattice energy (15 meV) of crystalline Bi are comparable. We believe that the underlying
mechanism for superconductivity in c-Bi is entirely different from that of a-Bi. The mechanism for
superconductivity in the latter is due to electron-phonon interaction which is firmly established.

Let us now look at the other models which suggest unconventional mechanisms
\cite{koley2017a, baskaran2017a, ruhman2017a, tewari2018a, ekkehard2018a}  for the
observation of superconductivity in Bi. S. Koley \textit{et.al.} \cite{koley2017a} suggested that fluctuating excitons could be the glue for superconductivity in Bi. They showed that a two-fluid model composed of preformed and dynamically fluctuating excitons coupled to a tiny number of carriers can  provide a unified
understanding of the anomalous temperature dependence of the resistivity below 1 K as well as superconductivity in Bi below 1 mK.
Our unpublished resistivity data (see fig.5 in the SM) clearly show deviation from T$^2$ dependence of the resistivity below 800 mK and ultimately
shows a T$^5$ behaviour down to 15 mK (in agreement with an earlier study \cite{Uher1977a}). Koley's model has
proposed that resonant scattering involving a very low density of renormalized carriers and the excitonic
liquid drives logarithmic enhancement of vertex corrections, boosting superconductivity in Bi. The model
also explains the temperature dependence of the normal state resistivity down to 15 mK.
However, at present it is not clear whether such a model can explain the anisotropy observed in the critical fields.
Bhaskaran \cite{baskaran2017a} suggested that a potential high $\mathrm{T_C}$ SC from the RVB mechanism is lost in
Bi. However, some superconducting fluctuations survive and the tiny Fermi pockets are
viewed as remnant evanescent Bogoliubov quasiparticles that are responsible for the anomalous normal state. Multi-band
character admits possibility of PT violating chiral singlet superconductivity Bi. It remains to be
seen whether such a chiral superconductor exists and can explain the observed anisotropy. Moreover,
Bhaskaran's model also suggests elements Sb and As should also exhibit SC at low temperatures. We did not observe SC
in single crystals of both of Sb and As (unpublished) down to 0.1 mK in a field of $0.4 \mu$T.
Ruhman and Lee \cite{ruhman2017a} have argued that conventional electron-phonon coupling is too weak to be responsible
for the binding of electrons into Cooper pairs. They showed that Bi is the first material to exhibit superconductivity
driven by the retardation effects of Coulomb repulsion alone. They claimed that SC of Bi at low carrier concentration arises only due to the long-ranged interactions that are capable of causing such an instability. In the absence of
any experimental evidence of a critical point, they investigated the more likely scenario in which the dynamically screened
Coulomb repulsion gives rise to an effective retarded attraction on the energy scale of the longitudinal plasma oscillations. Further, they
used an approximate isotropic band structure and the random phase approximation for the screened Coulomb interaction.
Within these approximations, they found the above-mentioned weak coupling instability. The transition temperature is greatly enhanced by the existence of a heavy hole band which has a large mass that allows for an enhancement of the static screening (Thomas-Fermi)
without enhancing the plasma frequency. They also showed that $\mathrm{T_C}$ is not dramatically decreased when the acoustic plasma
mode is absent. Therefore it was concluded that acoustic plasmon does not contribute to attractive interactions in Bi. They emphasized that the model works only in the s-wave coupling and the scenario might change
with higher angular momentum coupling that might result in the SC of Bismuth. Our analysis suggests
that electrons (not holes) are responsible for SC in Bi and that needs to be reconciled with \cite{ruhman2017a}.

Tewari and Kapoor \cite{tewari2018a} have suggested that superconductivity of Bi arises due to the electrons belonging to the three pockets.
They claimed that the electrons in Bi behave like a rare gas with inter-particle separation of nearly $185 \AA$.  In such a dilute system, the peculiar oscillatory behavior of the generalized electronic dielectric function at large distances can give rise to an attractive
interaction between two electrons. This model is valid at extremely low temperatures for a very low density electron gas, so that there is no electron–phonon interaction and the interaction amongst the electrons can be expressed in terms of weak two-body potential characterized by negative scattering length. The dilute nature of the electron gas is crucial and necessary requirement (i.e. $\mathrm{k_F \times a \ll 1}$, where a is the distance between atoms) to observe SC in Bi.  We have calculated the Sommerfeld coefficient($\gamma$) using the formula $\mathrm{\gamma=\frac{\pi^2}{3}k_B^2g(E_F)}$, where $\mathrm{g(E_F)}$ is the density of carriers at the Fermi level. Here the carrier comprises both electrons and holes. A simple estimate of $\gamma$, assuming the density of states (per pocket per spin) of electrons at the Fermi level, $\mathrm{N_0^e=9.2\times 10^{18}eV^{-1}cm^{-3}}$ \cite{ruhman2017a}, and the density of states (per spin) of holes at the Fermi level,  $\mathrm{N_0^h=3.5\times 10^{19}eV^{-1}cm^{-3}}$ \cite{ruhman2017a}, gives $\gamma \approx 4 \,\mu$J K$^{-2}$ mol$^{-1}$. This estimated value is quite close to the experimentally obtained $\gamma \approx 5 \,\mu$J K$^{-2}$ mol$^{-1}$ \cite{prakash2017a}. Therefore, a refinement of the model proposed by \cite{tewari2018a} is required to understand the temperature dependence of the critical field.
Finally, the model proposed by Ekkehard Krüger \cite{ekkehard2018a} relies on the hypothesis that the non-adiabatic Heisenberg model presents a mechanism
of Cooper pair formation generated by the strongly correlated atomic-like motion of the electrons
in narrow, roughly half-filled “superconducting bands” of special symmetry. In this case, the formation of
Cooper pairs is not only the result of an attractive electron–electron interaction but is additionally the
outcome of quantum mechanical constraining forces. According to his assertion  only these constraining forces operating in superconducting bands may produce eigenstates
in which the electrons form Cooper pairs. He argued that both Bi at atmospheric pressure and Bi at 122 Gpa
have nearly half-filled narrow electronic bands which may be responsible for superconductivity
in both these phases of Bi. However, this model does not provide any information on
superconducting properties of Bi apart from the estimation of $\mathrm{T_C}$. It is clear that more theoretical
work is required to understand the superconducting properties of this fragile electronic SC
of Bi.

\subsection*{Acknowledgment}
R.S. and A.P. acknowledge use of computational facilities at Dept. of Theoretical Physics. R.S. acknowledges support from the Department of Atomic Energy, Government of India, under Project Identification No. RTI 4002. R.S. thanks Pratap Raychaudhuri, Unmesh Ghorai and Mursalin Islam for useful discussions.

\bibliography{Bismuth.bib}


\end{document}